\documentclass[12pt]{article}
 
\usepackage{amsfonts}

\newcommand{\be}{\begin{equation}}
\newcommand{\ee}{\end{equation}}
 
\newcommand{\bea}{\begin{eqnarray}}
\newcommand{\eea}{\end{eqnarray}}

\usepackage{amsfonts} 
\usepackage[dvips]{graphicx}
\usepackage{amssymb} 
\usepackage{amsbsy}
\usepackage{hyperref}
\usepackage{units}
\usepackage{xcolor}
\usepackage{lscape}
\usepackage{csquotes}
\usepackage{booktabs}
\usepackage[normalem]{ulem}
\usepackage[round]{natbib} 
\usepackage{footmisc}

\begin{document}

\centerline{\bf
The effects of anger on automated long-term-spectra based speaker-identification}



\hfil

{\footnotesize 

\centerline{
Diana Valverde-M\'endez$^{a,b}$, 
Manuel Ortega-Rodr\1guez$^{a,}$\footnote{Corresponding author at: 
Escuela de F\1sica,  
Universidad de Costa Rica, 11501-2060 San Jos\'e, Costa Rica. 
E-mail addresses: 
dsmendez@princeton.edu (D.~Valverde-M\'endez), 
manuel.ortega@ucr.ac.cr (M.~Ortega-Rodr\1guez), 
hugo.solis@ucr.ac.cr (H.~Sol\1s-S\'anchez), 
avenegasli@ucdavis.edu (A.~Venegas-Li).}, Hugo~Sol\1s-S\'anchez$^{a}$, 
Ariadna~Venegas-Li$^{a,c}$}

\centerline{
$^a$ Escuela de F\1sica and 
Centro de Investigaciones Geof\1sicas,  
Universidad de Costa Rica, 11501-2060 San Jos\'e, Costa Rica}
\centerline{
$^b$ Department of Physics, Princeton University, Princeton, NJ 08544-0708, USA
}
\centerline{
$^c$ Physics Department, University of California at Davis, One Shields Avenue, Davis, CA 95616, USA}}

\hfil

\begin{abstract}
\noindent
Forensic speaker identification has traditionally considered 
approaches based on long term spectra 
analysis as especially robust, given that
they work well for short recordings, are 
not sensitive to changes in the intensity of the sample, and continue to function in the presence of noise and limited passband. 
We find, however, that anger induces a significant distortion of the acoustic signal for long term spectra analysis  purposes. 
Even moderate anger offsets speaker identification results
by 33\% in the direction of 
a different speaker altogether. 
Thus, caution should be exercised when
applying this tool.

\end{abstract}


\hfil

\section{Introduction}
The purpose of this article is to quantitatively determine the effects of emotional distortions (in particular, those of anger) on the long term spectra (LTS) analysis used for speaker identification (SPID). (For authoritative SPID review articles, 
see e.g. \citealt{hollien2013barriers, hollien2016approach}.) This is performed through a careful and replicable methodology.  

The objective of the SPID process is to identify an unknown speaker through voice analysis, usually under conditions which are not ideal. \textcolor{black}{The process often involves comparing one unknown voice against the voice of a known speaker (and thus, emotional state differences between recordings could be problematic).}
One of the fundamental challenges that SPID faces is the determination of  whether or not intraspeaker variability is smaller 
than interspeaker variability, and of how this relationship holds  
for different conditions \citep{hollien2002forensic}.  These conditions include distortions from varied sources, such as technological distortions due to the equipment used for the recordings, and environmental distortions caused by noise or harsh sounds in the background.  In particular, speakers may often be the source of distortion themselves, as a variety of feelings including fear, anger, and anxiety may be present 
(a situation likely to occur, for example, in forensic SPID when the speaker might be committing a crime).  These emotions trigger a 
change in speech production that reveals itself 
as a shift in the values of the measured signal parameters 
(such as frequencies and speech speed) \citep{williams1972emotions,  banse1996acoustic, johnstone2001effect}. 
As voice production consists of air pulses caused by the vibration of the vocal folds (which are then modified by the supralaryngeal vocal tract), the dominant factors of vocalization are the respiration patterns and the varying tension of the muscles involved in the process, \textcolor{black}{and these will most likely modify both the fundamental frequency produced and the position of the formants.}  Since respiration and muscle tone are highly correlated to emotions, it is therefore very likely that these changes in tension are 
detectable in the acoustic sound wave \citep{scherer1986voice}.

This topic, however, has not been extensively studied largely because there are ethical and methodological constraints that prevent the induction of 
controlled high magnitude emotions.
Current consensus \citep{johnstone2001effect} is that 
standard controlled laboratory conditions are feasible only for
low-intensity emotional responses,
for which it is difficult to
observe appreciable changes
in SPID effectiveness.  Another difficulty is how to induce the correct kind of emotion. 
\citet{martin1990induction} gives an overview on some of the possible techniques used to induce emotions, including emotional music, pictures, and self-generated techniques such as imagination and memories.  These methods are classified according to how the emotions are produced.  For the purpose of this article, the autobiographical recall approach was selected.  In this technique,  
subjects are asked to remember mood-evoking events in order to generate the desired emotion.  While this is not the only method applicable to the problem, it was selected for its simplicity and because it allowed the subjects to have privacy while recording. Furthermore, experiment blindness was assured by having the participants themselves assessing their level of anger, as described below.  

The remainder of this section discusses the workings of the LTS SPID process.  Although there are many different markers (also called ``vectors'') that help to distinguish two recordings from different speakers, ever since the pioneering work of \citet{hollien1977speaker}, one of the most common methods used in SPID is LTS analysis, also known as long term average 


\noindent
speech spectra (LTASS) analysis \citep{kinnunen2006use, ortega2015Cigefi}. 
The LTS analysis evinces the 
time-averaged quality (timbre) of the voice, an acoustic feature that allows a person to distinguish between, for example, 
a clarinet and a violin playing the same note (frequency) with the same intensity. The distribution is obtained computing the Fourier transform of the signal over a long period of time, say, 30 seconds; the idea is that the speaker has enough time to move through all the sound phase space.

This vector has been widely studied in terms of how efficiently identification is performed and has been found to be one of the most reliable ones, mainly because it continues to function in the presence of noise
and limited passband  \citep{hollien2002forensic}. 

One of the challenges of using this vector is how to define the correlation between two spectra. There are several approaches to this problem, including assigning a specific number to each LTS set (according to some algorithm), or even visual inspection of the plots.  For the purposes of this article, a more sophisticated approach was required,  \textcolor{black}{as the sought differences turned out to be subtle}.  Two correlation coefficients were considered for this purpose:  the Standard Deviation of the Differences Distribution (SDDD) \citep{harmegnies1988sddd} and the Bravais-Pearson cross-correlation coefficient, $R$ \citep{stanton2001galton}. Exploratory experiments performed 
by our group without the anger component \citep{ortega2015Cigefi} showed that the Bravais-Pearson correlation coefficient gives the best results for this line of research and was therefore chosen.  

In the Bravais-Pearson approach, 
a spectrum in the LTS analysis is considered to be 
a $k$-dimensional vector, with $k$ frequency channels.  Thus, the spectrum may be defined as
\be
S \equiv (S_1, \dots , S_i, \dots, S_k) \, ,
\ee
where $S_i$ is the level of the $i$-th frequency component (Harmegnies, 1988).  In this context, the $R$ coefficient measures how related two LTS samples  are. 
$R$ is defined as: 
\be
R_{SS^\prime} \equiv \frac{1}{k} \, \frac{\sum\limits_{i=1}^{k}(S_i - M_S)
(S^\prime_i - M_{S^\prime})}{\sigma_S \, \sigma_{S^\prime}}  \, , 
\ee
where $M$ stands for the mean of each spectrum, and $\sigma_S, \sigma_{S'}$  
are the respective standard deviations.  The Bravais-Pearson coefficient provides several advantages. 
Not only does it have a high discriminating ability, but also it is independent of the overall differences in the intensities of the two spectra \citep{harmegnies1988sddd}.  This allows for comparison of recordings that were conducted  under different microphone placement or environmental conditions.

Most of the studies on the relationship between speech and emotions have 
focused on the ability to distinguish between different emotional states of a speaker 
\textcolor{black}{(e.g. anger, sorrow, fear)}
through signal analysis 
\citep{williams1972emotions, fuller1972detection, scherer1986voice, johnstone2001effect, harnsberger2009stress}.
In particular, LTS analysis has been used to attempt to identify emotional states and depression \citep{pittam1987long}, sometimes making use however of human filtering \textcolor{black}{(perception tests) in the 
research} process   \citep{banse1996acoustic}.  

Less common is research on how emotions or disguise affect SPID. 
\citet{rodman2000computer} 
recommend and plan out research to
study the effects of disguise on SPID, although they do not carry it out.
More related to our article,
\citet{hollien1977speaker} studied the effects of stress 
(induced via electroshocks on the subjects)
on the LTS of a signal, 
but they did not find a significant deviation from adequate SPID. Other than this, to the best of the 
authors' knowledge, there is no 
study on the effect of emotions on LTS SPID.

 

\section{Materials and methods}

\subsection{Recorded subjects}
The SPID problem has many variables. Consider, for example, the
gender, geographical origin, and age of the subjects \citep{hertrich1987sexual, linville2002source, hollien1977speaker, pittam1987long, yuksel2017long}. 
For this reason, \textcolor{black}{and in order to work in a smaller phase space in this exploratory endeavor,} subjects with similar characteristics were chosen, in order to highlight the effects of emotion and reduce the overall complexity of the problem.
The subjects met the following restrictions: they are males
in the age range 18--25, and  
their geographical origin is 
Costa Rica's Central Valley Metropolitan Area (this region consists of the four most populous cities of the country, located in the country's 
densely populated mid-region).  
Subjects also attended primary school in this region.  
\textcolor{black}{Variations in speech would presumably be 
less marked under these conditions.}

\subsection{Recording conditions}

Recording conditions were standardized to ensure homogeneous results.
Each of the speech samples was collected in a private location 
\textcolor{black}{(which was not, however, the same for all participants)} where the subjects would feel comfortable.
Additionally,
the speech was required to be fluid and not scripted, and top of the range smartphone microphones were used,
such as the iPhone and Samsung Galaxy series. 

According to the emotional state, there were two types of recordings:
{\it Normal recordings}, in which  subjects were asked to speak about their  day-to-day life, 
in order to have as little emotional response to the topic as possible;
{\it angry recordings}, in which subjects were asked 
\textcolor{black}{to invoke an angry state by means
of describing a situation that made them angry.}  
The interviewer was not present in order to avoid inhibiting the subjects.

The sample length was 45 and 60 seconds for the normal and angry 
cases, respectively (extra time for the angry cases allowed for some transition
time \textcolor{black}{for anger to presumably build up}). 

\subsection{Methodology}

32 subjects who met the selection criteria described in Section 2.2 were interviewed.
This number was chosen  
so as to have a significant sample \citep{nist} of the Costa Rican 
Central Valley Metropolitan Area population.  
Each interview consisted of three recordings.
For 16 of the 32 interviews, the following recording order 
was used: normal-normal-angry.
For the other 16 interviews, we used instead a
normal-angry-normal order.
The motivation behind having two types of order was to
reduce the effects of possible systematic errors due to 
sampling order.

The subjects were taken to a private place where a script was read to them. 
The interviewers explained that this was a research project for the Universidad de Costa Rica and that the contents of the recordings would not be used or heard.   
It was emphasized that only the acoustic parameters of the samples were to be used and that the level of anger would be only determined by the subjects' self-rating at the end of the recording process (a characteristic that also provided blindness to the methodology).  Furthermore, the nature of the project was described only in general terms to avoid affecting the natural outcome of the recordings.

In the case of normal recordings, the speakers were instructed to talk for 45 seconds about their life (for example, their day, their pet, a recent event, or their interests). 
In the case of anger recordings, the subjects were asked to 
\textcolor{black}{invoke a state of anger by} speaking for one  minute about something that made them angry, for example, a person they hated or some event that especially bothered them.  The subjects were instructed 
not \textcolor{black}{to fake the emotion (for example, by forcing it with 
shouts).} They were left alone with the recorder and were told to speak whenever they felt ready.
   
Once the anger recording was completed, the participants were asked to rank the anger
\textcolor{black}{they had during the recording} in a scale   
from 1 to 5, where 1 meant ``not able to get angry,'' 
3 meant ``moderately angry,'' and 5 meant ``furious.''

Note that
when the first order was used, subjects did not know about the 
emotional-response part until they had to do the anger recording, 
whereas, in the case of the normal-angry-normal order, the speaker knew about this aspect of the research during his last normal recording. As mentioned, 
both orders were used and averaged so as to reduce order bias.

\section{Data processing}
 
The processing in which the present research is based was developed by our group \citep{ortega2015Cigefi} for 
SPID (without the anger component)  and has been extensively tested and optimized for best identification results. 

The data processing corresponding to the present paper can be summarized as follows. 
Thirty-second segments were obtained from the original recordings. In the case of the anger samples, this was particularly important to filter out 
the possible transition from normal to angry state. 
The Audio processing software Audacity 2.1.0 \citep{team2015audacity}
was used to obtain the Fast Fourier Transform (FFT) for each recording. 
A Hanning window was used with 4096 frequencies sampled. This parameter was
empirically determined to render the best results in our previous exploratory studies.
Each FFT was stored as a text file for further processing.

Next, a  C++ code computed the 
Bravais-Pearson correlation coefficient for the samples.  
Finally, the average, the sample standard deviation (SD) 
and the standard error of the mean (SE) for the correlation coefficients were obtained for the cases of normal-normal-angry and normal-angry-normal recordings.

\section{Results}
 
Table \ref{table1} displays the 
results of the process
described in the previous section. 
\begin{table}[b]
   \centering
     \begin{tabular}{@{} lll @{}} 
      \toprule
           & Normal-normal & Normal-angry\\
      \midrule
      Average      & 0.950 & 0.934 \\
      Standard deviation       & 0.028  & 0.037 \\
      Standard error of the mean       & 0.005  & 0.005 \\
      \bottomrule
   \end{tabular}
   \caption{Statistics of the intra-speaker
            Bravais-Pearson correlation coefficient $R$. A total of 32 subjects were
            used for the normal-normal and normal-angry contrast.}
   \label{table1}
\end{table}
32 subjects were interviewed and the Bravais-Pearson correlation coefficient $R$ was calculated to obtain the normal-to-normal correlation and 
the normal-to-anger correlation. There is a noticeable difference between the two obtained averages. To understand how significant this difference actually is, it is useful to compare it to the results of the aforementioned research  performed by our group  \citep{ortega2015Cigefi}, 
which is a simplified version of the process described in the present paper
(as the anger element was absent),
although the experimental conditions
were the same. 
There, the average correlation coefficient $R$ between two different speakers 
was found to be 0.890 with an SE of 0.010, whereas  
same-speaker correlation had a mean $R$ coefficient of 0.955 with an SE of 0.005.
 
Comparing these three correlations:  
0.950 (normal-normal, same speaker),
0.934 (normal-angry, same speaker), and 
0.890 (different speakers, both normal),
we see that the effects of anger deviate the signal (from
what would be otherwise expected) 
a significant 33\% of the difference in correlation between the same speaker under normal circumstances and two different speakers under normal circumstances.
This is remarkable since the average self-reported anger was 2.9 in the 1-5 scale described in Section 2.3, meaning that on average the subjects were
only moderately angry. The distribution of the level of anger for our subjects is presented in Table \ref{table2}

\begin{table}[htb]
   \centering
     \begin{tabular}{@{} ll @{}} 
      \toprule
           Number of subjects& Self-reported anger level\\
      \midrule
      5       & 4 \\
      18       & 3 \\
      9      & 2 \\
      \midrule
      Average & 2.9\\
      Standard deviation & 0.65\\
      Standard error of the mean & 0.12\\
      \bottomrule
    \end{tabular}
   \caption{Self-reported anger level for subjects; 
   1 meant ``not able to get angry,'' 
   3 meant ``moderately angry,'' and 5 meant ``furious.''
   }
   \label{table2}
\end{table}

Higher anger ratings were expected to generate larger effects on the LTS SPID. To test this hypothesis, we selected only the five angriest subjects (those who self-rated 4 on the 1--5 scale).
Table \ref{table3} shows the results obtained for this case. For strong anger, the deviation from normality is close to 50\% of the shift
to a different speaker.
This shows that the effects of anger on the LTS SPID do indeed grow as anger increases.

\begin{table}[htbp]
   \centering
     \begin{tabular}{@{} lll @{}} 
      \toprule
           & Normal-normal & Normal-angry\\
      \midrule
      Average      & 0.950 & 0.922 \\
      Standard deviation       & 0.030  & 0.047 \\
      Standard error of the mean       & 0.013  & 0.015 \\
      \bottomrule
   \end{tabular}
   \caption{Statistics of the intra-speaker
            Bravais-Pearson correlation coefficient for the (self-rated) five angriest recordings. As expected, strong anger has a larger effect on 
            LTS SPID.}
   \label{table3}
\end{table}


\section{Conclusions}

Even though some authors praise LTS SPID for being robust to speaker stress \citep{hollien1977speaker}, 
we have found that there is, in fact, a significant distortion
on the human voice due to anger for LTS SPID considerations. 
Even as the emotional response obtained from the participants was only moderate, 
we found an appreciable difference in the 
correlation coefficients between the cases of 
normal-normal recordings and normal-anger recordings. 
Moderate anger offsets SPID results by a significant 33\% in the direction of 
a different speaker. 
Furthermore, these results were obtained with a method which can 
be fully automatized, providing an objective approach independent of human perception errors. 
\textcolor{black}{The method also avoids assessing the sincerity of the participants, 
and is therefore in agreement with the code of practice of the 
International Association for Forensic Phonetics and Acoustics \citep{iafpa}.}

Our results are relevant for forensic research as LTS analysis has been traditionally considered a robust 
vector in SPID, especially since it is not sensitive to changes in the intensity of 
the speech sample, works well for short recordings, and
continues to function in the presence of noise
and limited passband. 
The results of the present paper, however, indicate that one should be cautious
when using LTS SPID 
to calculate likelihood ratios 
in a scenario of anger, even if this anger is not strong. 

As other emotions might also affect significantly the effectiveness of LTS SPID, 
they are also worthy of future 
automated study. The study could
also be performed on women, or
on speakers of a different language.

\section*{Acknowledgments}

This work was supported by
grant 805-B2-175 of the Universidad de Costa Rica's Vicerrector\1a de 
Investigaci\'on and the CIGEFI.
\textcolor{black}{The authors 
would also like to acknowledge the useful comments of  
two Speech Communication anonymous reviewers.}
 

\bibliographystyle{apalike} 
\bibliography{ltsAnger}









\end{document}